# Physical origins of protein superfamilies

Konstantin B. Zeldovich, Igor N. Berezovsky, Eugene I. Shakhnovich*

*Corresponding author

*Department of Chemistry and Chemical Biology, Harvard University, 12 Oxford St,*

*Cambrigde, MA 02138*

**Abstract**

In this work, we discovered a fundamental connection between selection for protein stability and emergence of preferred structures of proteins. Using standard exact 3-dimensional lattice model we evolve sequences starting from random ones and determining exact native structure after each mutation. Acceptance of mutations is biased to select for stable proteins. We found that certain structures, "wonderfolds", are independently discovered numerous times as native states of stable proteins in many unrelated runs of selection. Diversity of sequences that fold into wonderfold structures gives rise to superfamilies, i.e. sets of dissimilar sequences that fold into the same or very similar structures. Wonderfolds appear to be the most designable structures out of complete set of compact lattice proteins. Furthermore, proteins having wondefolds as their native structure tend to be most thermostable among all evolved proteins. This effect is purely due to the favorable geometric properties of wonderfolds and, thus, dominates any dependence on sequences. The present work establishes a model of *prebiotic* structure selection, which identifies dominant structural patterns emerging upon optimization of proteins for survival in hot environment. Convergently discovered prebiotic initial superfamilies with ''wonderfold'' structures could have served as a seed for subsequent biological evolution involving gene duplications and divergence.

**Introduction**

One of the most striking discoveries in structural biology is highly uneven distribution of protein fold usage (1-5)The ability of protein structures to accommodate many unrelated sequences was demonstrated in theory and simulations and is generally understood on theoretical grounds (6-9). At the same time, the observation that only certain protein structures form superfamilies while others do not and that the size of superfamilies varies greatly, represents one of the major puzzles in structural biology. Revealing the cause of such unequal distribution is a key to our understanding of protein evolution from structural perspective. Two fundamentally distinct explanations can be hypothesized. First, it is possible that modern highly populated folds were selected by chance early in evolution and their dominance was preserved in the process of divergent evolution that resulted in modern protein universe (10). The second possibility is that most abundant folds have certain intrinsic advantage, i.e. they emerged as a result of some form of purifying selection. Finkelstein and coworkers suggested that folds corresponding to populated superfamilies are highly designable, i.e. they can accommodate a large number of sequences (11). Subsequent study by Li and coworkers (12) demonstrated, using an exact protein model of 27-mer heteropolymer on the cubic lattice (13) that indeed some structures can adopt more sequences than others, i.e. they appear to be more designable. The applicability of this seminal finding to real proteins was uncertain since no transferable structural determinant of designability was found at that time. More recently, England and Shakhnovich (14) found the structural determinant of protein designability using an analogy between protein design and statistical mechanics of spin models. They showed that so-called contact traces (traces of powers of a protein's contact matrix), well-approximated by powers of maximal eigenvalue $\lambda_{max}$ of the contact matrices, serve as highly significant predictors of protein designability (14).. This property is directly transferable to real proteins since it is straightforward to calculate their contact matrices. Indeed, in recent study a statistically significant correlation between the structural determinant of designability and the size of gene families was found (15).. However the explanation of uneven fold population based on the concept of designability remained inconclusive because no realistic selection mechanism that results in emergence of preferred most designable structures was found despite previous efforts (16-19). Here, we show that the requirement to maintain high thermostability results in emergence of superfamilies of most designable folds. This finding provides the missing link between modern physical views on protein structure fitness (designability) and possible prebiotic mechanisms that gave rise to uneven fold usage and emergence of superfamilies *en route* of protein evolution.

Our approach is based on the 27-mer lattice model of protein whose all compact conformations were enumerated (13), providing the basis for exact statistical-mechanical analysis. Lattice models of proteins, including compact 27-mers proved extremely useful in the past as they helped to provide crucial insights into protein folding (20-24), protein design and selection (7, 25, 26), and protein evolution(17, 18, 27, 28). Here we study the simplest model of convergent prebiotic evolution whereby we carry out the simultaneous search in sequence and structure spaces to find stable model proteins. This approach is an extension of the earlier works (6, 29-31) where a design procedure based on stochastic optimization in sequence space with *fixed* target structure has been developed. Here we no longer require that the structure is fixed: a new native conformation (the one of lowest energy among all 103346 compact folds) is determined after each attempted "move" in sequence space. This crucial extension allows us to explore the issue of whether there are preferred native conformations of stable proteins and if so, what are their properties.

In this microscopic model of physical selection the search starts from proteins with random sequences and results in convergent repetitive discovery of proteins having special specific structures, "wonderfolds". Multiple independent runs of sequence design algorithm yield many unrelated sequences folding into the same structure, i.e. prototypic protein superfamilies. In contemporary proteins superfamilies (2) are the sets of proteins with apparently non-homologous sequences (i.e. undetectable by sequence alignment programs such as BLAST) yet similar structures (8). To link the results of our simulations with physically measurably quantities, we demonstrate that our energy-based design procedure selects model proteins with increased thermodynamic stability (melting temperature). The present work establishes a model of *prebiotic* structure selection, which identifies dominant structural patterns emerging upon optimization of model proteins for survival in hot environment.

**Materials and methods**

**Model.** Our lattice model of protein structure consists of the complete set of 103346 compact 27-mers on the 3x3x3 cubic lattice (13). The energy of a given conformation $k$ is calculated using the Miyazawa-Jernigan contact energy matrix $\varepsilon_{ij}$ (32):

$$E_k = \frac{1}{2} \sum_{i,j=1}^{27} \varepsilon_{Si,Sj} C_{ij}^{(k)},$$

where $C_{ij}^{(k)}$ is the contact matrix of conformation $k$, and the sequence is represented by the vector $S_i$, whose components correspond to the type of the amino acid in position $i$ along the chain. We perform all the design simulations in the framework of two different models of amino acid content, random amino acid composition with equal probabilities of all residues, and a random gene model, where all nucleotides have equal probability. The results for the random amino acid composition are presented in the paper; while the results for the random gene model are presented in Supporting Information. As the complete enumeration of all sequences is infeasible, we resort to sequence design via a Monte Carlo procedure. For each initial sequence, we are minimizing its ground state energy by attempting to permute two arbitrarily chosen amino acid residues, so that the amino acid composition is preserved in a run. We then find the ground state by selecting the conformation with lowest energy out of all 103346 compact structures. If the ground state energy decreases after permutation, the new sequence is accepted, otherwise, the standard Metropolis scheme with "selective temperature" $T_{sel}=0.1$ is applied. Throughout the simulation, we monitor the ground state energy, and whenever it dips below the currently achieved minimum, we record the sequence, its ground state energy, and ground state structure. Ground state energy minimization is stopped after 2000 Monte-Carlo steps, and the sequence corresponding to the lowest energy ever detected in the run is the result of the run of the minimization procedure. We checked (data not shown) that simulation runs longer than 2000 Monte-Carlo steps do not provide a significant improvement of the ground state energy. The sequences obtained as described above will be further called the evolved sequences. In contrast to a related algorithm proposed by Tiana et al.(33), we are able to perform the complete enumeration of compact structures for a given sequence, and thus establish its ground state exactly.

**Melting temperature.** To calculate the melting temperature of a model protein, we consider the Boltzmann probability $P$ of the protein being in its ground (native) state at temperature $T$,

$$P(T) = \frac{e^{-E_0/T}}{\sum_{i=0}^{103345} e^{-E_i/T}},$$

where $E_0$ and $E_i$ are the ground state energy and energy in $i$-th conformation, respectively. The melting temperature $T_m$ is then found from the condition $P(T_m)=0.5$.

**Results and Discussion**

We perform 20000 runs of the sequence design procedure starting each run from a randomly generated sequence, and find the native structures for each of the 20000 evolved sequences. Figure 1 shows the probability that a structure is a ground state for $k$ sequences out the total of 20000 evolved sequences. For comparison we provide on Figure 1 the same plot for random sequences. It turns out that some structures can adopt up to 19 evolved sequences (red curve), while we observed no cases when more than 5 random sequences had the same ground state structure (black curve). For random sequences, this probability should be compared with the Poisson distribution

$p(k)=e^{-\lambda}\lambda^k/k!$

with $\lambda=20000/103346$ (blue curve). The deviation of the distribution for random sequences from the Poisson distribution is due to variation in designability of the lattice structures first observed by Li et al. (12, 34). Here, we demonstrate that our procedure of sequence design efficiently uncovers special structures, wonderfolds, (Figure 1) that serve as native structure to unusually large number of independently discovered evolved sequences.

How similar are the sequences belonging to an emerged superfamily, i.e. evolved sequences that fold into the same native conformation? To characterize sequence identity, we use the Hamming distance between two sequences, defined as the number of positions in the two sequences where the amino acid residues are different. Figure 2 shows the distribution of Hamming distances for three sets of sequences within superfamilies corresponding to the three most populated wonderfolds (average Hamming distance 21.3, roughly corresponding to average sequence identity of less than 25%), and the distribution of Hamming distance between random sequences (average Hamming distance 25.7, sequence identity less than 10%). Lower Hamming distance between the evolved sequences within a superfamily, compared to random ones, reflects universal physical requirements of fold stabilization that all sequences folding into a given structure must satisfy. Interestingly, a similar criterion of sequence identity less than 25% was adopted by Holm and coworkers in their definition of structurally similar yet distant in sequence families in construction of DALI domains within FSSP database effort (35). While the sequence identity within emerged superfamilies is somewhat greater than that for random sequences it is certainly too low to be identifiable by standard sequence alignment tools, i.e. here we indeed observe the formation of superfamilies.

What makes model protein structures wonderfolds? To reveal a quantitative structural characteristic of advantageous folds, we consider the maximum eigenvalue ($\lambda_{max}$) of a structure's contact matrix, which serves as the structural determinant of its designability (14). We present the scatter plot for the number of evolved sequences $k$ that a structure adopted in the process of our sequence selection versus $\lambda_{max}$ of its contact matrix (Figure 3a). This plot reveals a very strong positive correlation between the $\lambda_{max}$ of a structure and the maximum number of sequences found in the process of selection that fold into that structure. This result establishes a crucial connection between designability and formation of superfamilies: wonderfolds, being the most designable structures, form superfamilies in the process of sequence selection that favors highly stable proteins. It also shows an interesting interplay between statistical factors and structure-related bias towards wonderfolds: While low-$k$ structures, those adopting only few sequences, can span a broad range of $\lambda_{max}$, the high-$k$ structures, forming highly populated superfamilies are exclusively wonderfodls with high $\lambda_{max}$. Due to this interplay between selection and chance, the scatter plots between designability of a fold (its $\lambda_{max}$) and its gene family size ($k$) may be rather broad as suggested by Figure 3a. However, if one bins the data in $\lambda_{max}$ bins and evaluates correlation between the logarithm of average $k$ in a bin and $\lambda_{max}$ it would be extremely strong, with correlation coefficient R=0.98 (Figure 3b). If designability was not a factor, the graph in Figure 3b would have been a horizontal line at the constant level of ln (20000/103346) = -1.7. The pronounced slope of the graph is a very clear illustration of the role of designability, where some (high-lambda) structures are more populated than others. It also serves as a direct demonstration that $\lambda_{max}$ is the structural determinant of designability. Exactly the same effect - broad triangle-shaped scatter plot for raw data and very high correlation for binned data - is seen in the analysis of correlation between designability and gene family sizes in real data (15)

Given the high propensity of evolved sequences to populate wonderfolds, we can expect that wonderfolds (and their corresponding sequences) should manifest their favorable properties in terms of physically measurable quantities. For example, stability of a protein is related to its melting temperature $T_m$: the higher the melting temperature is, the more stable is the protein. Indeed, Figure 4 shows a drastic increase of the melting temperature of proteins having wonderfolds as their native structures as compared to model proteins with random sequences. The average melting temperature of 323 sequences populating the "top 20" wonderfolds is 1.10 (in dimensionless units corresponding to the Miyazawa-Jernigan amino acid interaction energies), whereas the average $T_m$ of random sequence/structure pairs is as low as 0.249. As noted above, the designability of a structure is determined by the $\lambda_{max}$ of its contact matrix. It turns out that the structures with a higher $\lambda_{max}$ yield statistically increased melting temperatures (Figure 5), which further corroborates the link between structural properties of wonderfolds and their stability.

Finally, the most direct way to demonstrate the inherently high stability of wonderfolds is a blind test comparing the melting temperature of proteins having wonderfolds as their native structures and proteins with randomly chosen structures with similarly designed sequences. For comparison with the 20 wonderfolds, we chose 20 random structures, and performed a standard fixed-structure sequence design procedure into these structures as described in (6, 29) starting in both cases from exactly the same random sequences. We designed 50 sequences for each of the 20 wonderfolds and 20 randomly chosen structures making 2000 Monte Carlo steps at the same selective temperature $T_{sel}$=0.1, repeating the condition of the original sequence selection described in this work. In case of random structures, we additionally performed longer and presumably more exhaustive design runs of 200000 Monte Carlo steps. The distribution of melting temperature of the resulting sequence/structure pairs is presented in Figure 6. Wonderfolds feature significantly higher melting temperatures (red, $<T_m>$=1.09) than randomly chosen structures (black), $<T_m>$=0.83 for short design runs and $<T_m>$=0.96 for long design runs (blue). Importantly, for random structures, even a much more exhaustive design (100-fold longer) does not reach the same level of stability as in the case of shorter design runs for wonderfolds. These results demonstrate that wonderfolds were especially advantageous when the search in sequence space was limited, i.e. at the earliest (prebiotic) stages of evolution. It seems therefore likely that emergence of wonderfolds dominated the physical selection and design of first thermostable proteins in the course of prebiotic evolution (36).

**Conclusions**

The discovery that protein superfamilies with wonderfold structures emerge convergently as a result of sequence/structure selection to optimize protein stability is the key finding of the present paper. For the first time we have explicitly and rigorously demonstrated in an *a priori* simulation the emergence of a small number of structural patterns that are inherently more favorable viz a viz a physical requirement of thermostability regardless of initial sequence from which sequence selection starts. Using a lattice model, we revealed that native state energy optimization naturally leads to the convergence towards a limited number of conformations with very specific structural properties, separating them from the bulk of compact 27-mers. Native energy optimization in our simulations, where aminoacid composition is fixed in each run, is apparently equivalent to optimization of stability: We searched for model proteins sequences having the lowest possible ground state energy, and found that evolved model proteins have a remarkably high thermostability. Thus, we show that prebiotic selection of thermostable structures can result in emergence of superfamilies around preferred most designable folds. Indeed, it has been convincingly demonstrated that proteomes from ancient archaea exhibit clear structural bias towards more designable folds (36) and, in particular, proteins from Last Universal Common Ancestor (LUCA) were found to be significantly more designable, than other, later evolved proteins (15). Also we note that Taverna and Goldstein (17) found, using a different, 2-dimensional, lattice model, that most designable structures are most robust in selection when requirement of even marginal stability is imposed – a result consistent with our findings.

As our model system mimics such basic common features of natural proteins as chain connectivity, compactness, unique sequence-structure relationship and highly complex energy landscape, one can expect similarity between presented convergent mechanisms of selection of 27-mers and the convergence of ancestral protein-like structures into *precursors* of contemporary protein folds at the stage of prebiotic evolution. Continuing the analogy, one can also infer that sequence selection mechanism simulated by our design procedure reproduces prebiotic events that resulted in the formation of sets of sequences favoring certain ancestral folds. Later on, these sets of sequences underwent subsequent mutations and natural selection via duplication and divergence *en route* of biological evolution, eventually leading to their divergence and emergence of modern protein universe with all its signatures such as modern sequence superfamilies (2, 37) and scale-free structural organization (10). Thus, the discovery of specialized structures, wonderfolds, among compact lattice 27-mers has important implications for our understanding of the physics of prebiotic selection mechanisms that presumably seeded subsequent biological evolution of natural proteins.

**Acknowledgments**

This work is supported by NIH. INB is a Merck Fellow.

**Figure Legends**

**Figure 1.**

Probability *p(k)* of finding *k* sequences that fold into the same structure from a pool of 20000 sequences . For random sequences (black curve), no structures populated by more than 5 sequences are observed. On the contrary, evolved sequences (red curve) tend to crowd specific structures, with up to 19 sequences in each of the most popular structures, and the distribution dramatically deviates from the null model of random population of folds in the form of Poisson distribution (blue curve). The value of *p(19)* for the Poisson distribution equals to $2.5 \cdot 10^{-14}$.

**Figure 2.**

Distribution of mutual Hamming distances between random sequences (black curve) and within superfamilies of sequences corresponding to the three most populated wonderfolds (red, green, and blue curves).

**Figure 3.**

**a.** Dependence of the number of evolved sequences corresponding to a structure on the maximum eigenvalue ($\lambda_{max}$) of the structure's contact matrix. One can clearly see that only most designable structures – wonderfolds – form large superfamilies containing many (up to 19) dissimilar sequences. A notable property of compact 27-mers is the gap in the distribution of the eigenvalues at $\lambda_{max} \approx 2.92$, which naturally separates wonderfolds from the bulk of the structures.

**b.** Dependence of the logarithm of the average number of evolved sequences k within a bin $\lambda_{max,i} <\lambda_{max}<\lambda_{max,i}+\Delta\lambda$, with $\lambda_{max,i}$ corresponding to 50 bins spanning the range from 2.52 to 2.96. All 103346 compact structures (including those that were not found in evolution simulations, i.e. having k=0) are considered here. A linear approximation (black line) is shown, correlation coefficient R=0.98.

**Figure 4.**

Distribution of melting temperatures of 20000 proteins with random sequences (black histogram) and 323 evolved sequences that fold into 20 most populated wonderfolds (red histogram). Evolved sequences melt at significantly higher temperatures than random ones.

**Figure 5.**

Correlation between the maximum eigenvalue $\lambda_{max}$ of the structures and the melting temperature $T_m$. Structures with evolved sequences (green), and especially the top 20 wonderfolds (red) correspond statistically to higher $\lambda_{max}$ and $T_m$ than random ones (black). For random sequences, the average $\lambda_{max}$ is 2.71 (standard deviation $\sigma=0.072$), and average $T_m$ is 0.258 ($\sigma=0.15$); for all evolved structures and top 20 wonderfolds the average $\lambda_{max}$ is 2.83 ($\sigma=0.055$) and 2.93 ($\sigma=0.016$), and average $T_m$ equals 0.881 ($\sigma=0.33$) and 1.307 ($\sigma=0.25$), respectively.

**Figure 6.**

Distribution of melting temperatures of 1000 sequences designed into the top 20 wonderfolds (red histogram), and into 20 randomly selected structures (short design runs for the black histogram and long design runs for the blue histogram, see in the text). In all cases, sequences designed into wonderfolds yield higher melting temperatures.

Figure 1

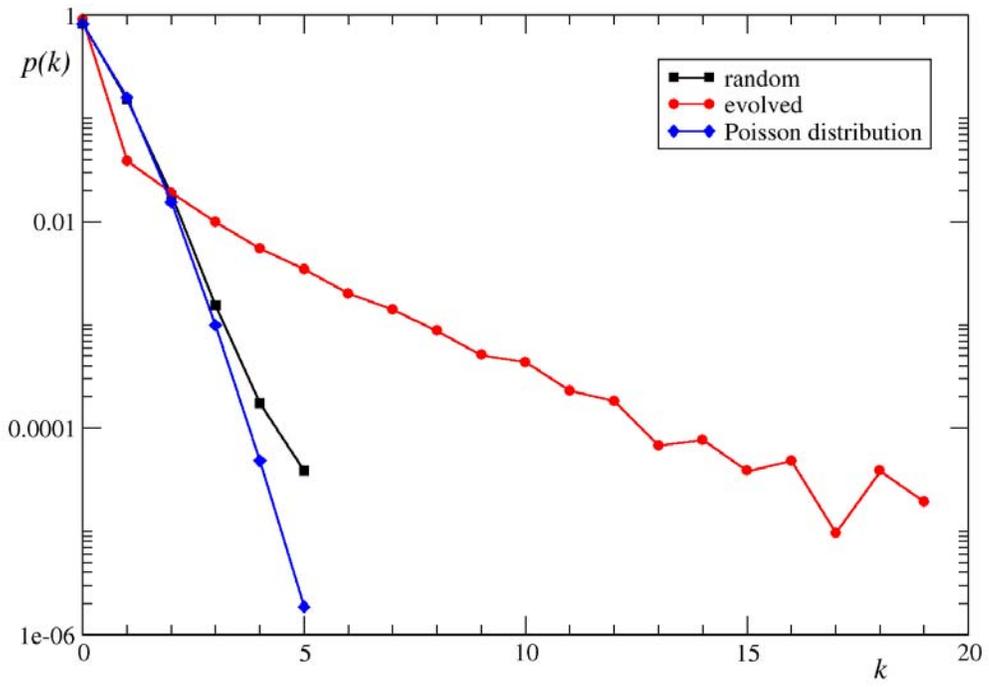

Figure 2

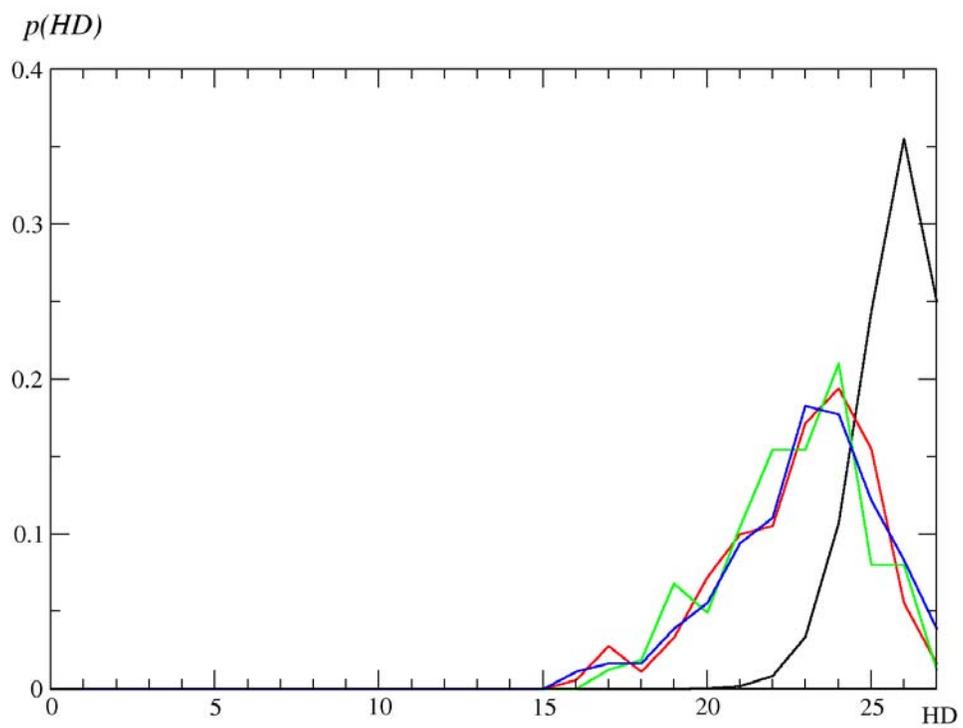

Figure 3a

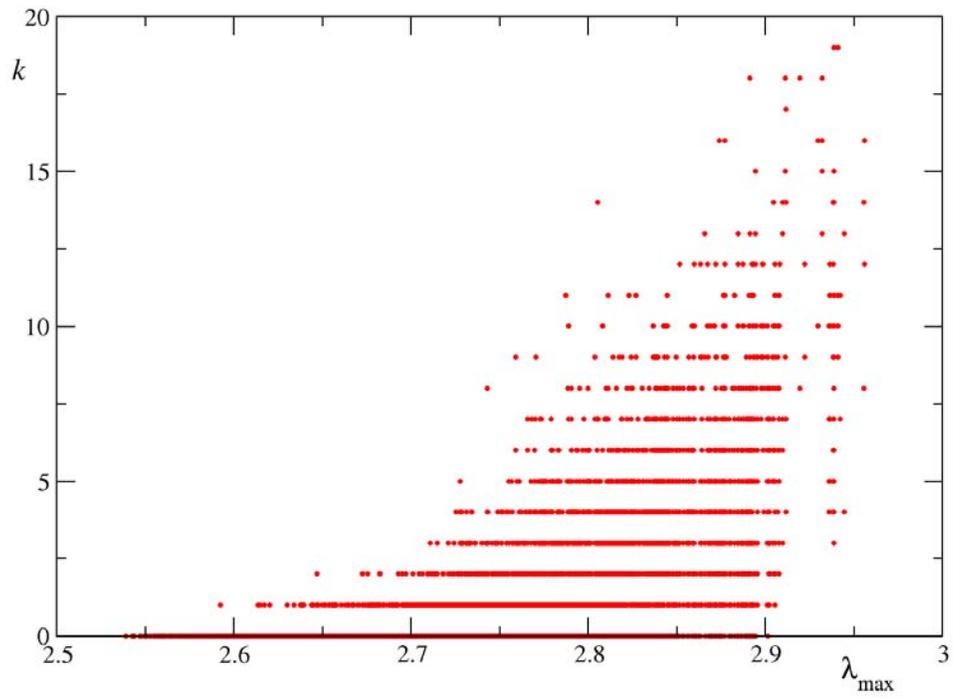

Figure 3b

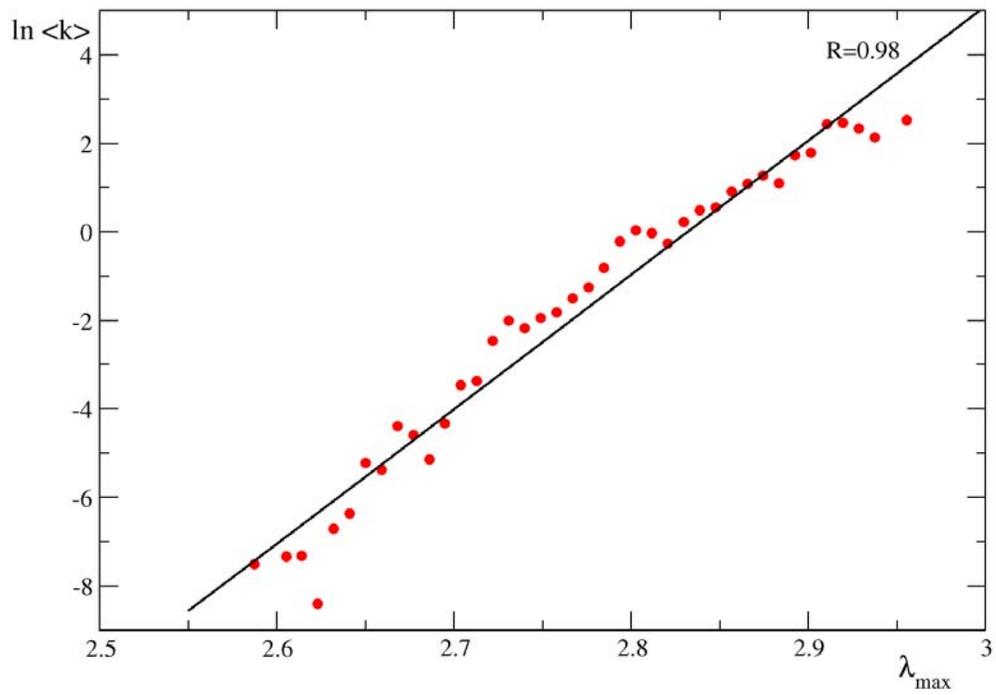

Figure 4

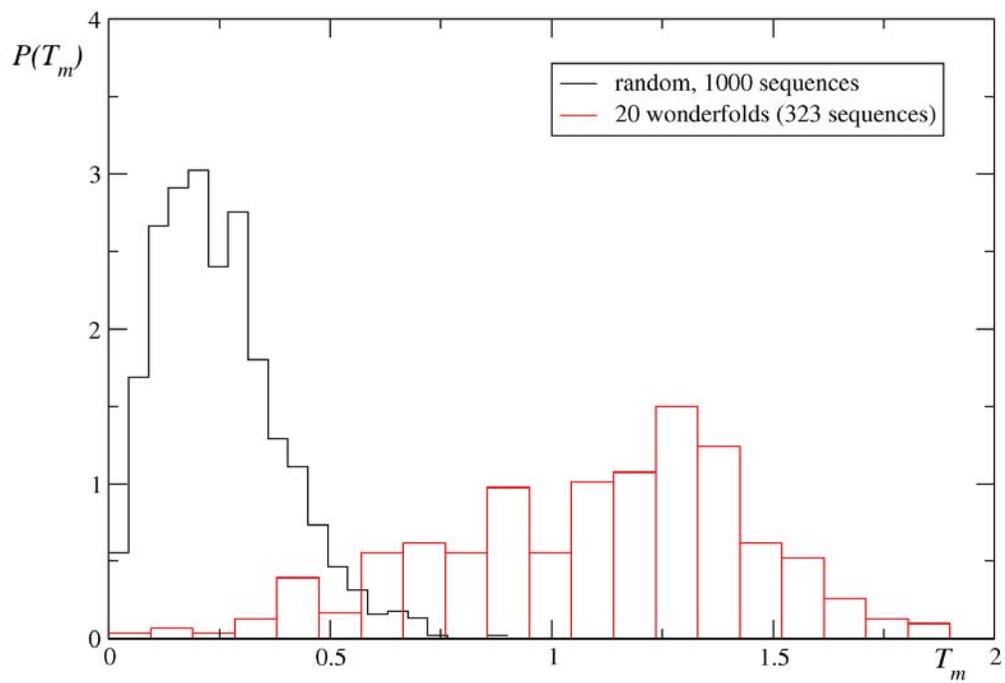

Figure 5

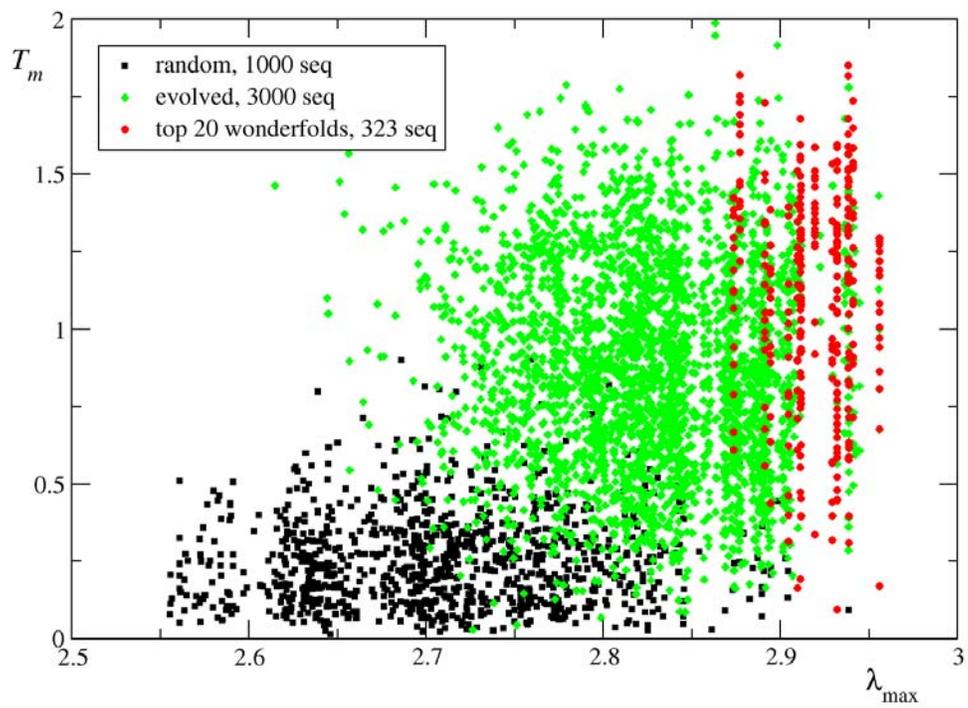

Figure 6

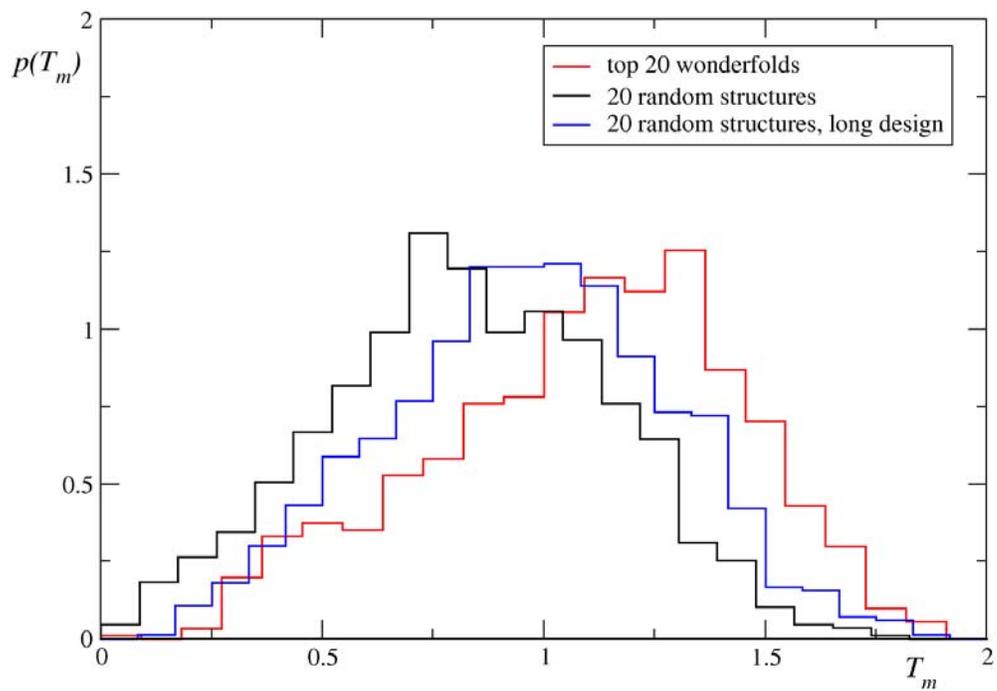

# Supporting Information

Here we present the results of the simulations similar to those described in the main text. The difference is that the amino acid composition of the sequences is determined by the random distribution of nucleotides in initial random sequences.

**Figure Legends**

**Figure S1.**

Probability $p(k)$ of finding $k$ sequences that fold into the same structure from a pool of 20000 sequences . For random sequences (black curve), no structures populated by more than 5 sequences are observed. On the contrary, evolved sequences (red curve) tend to crowd specific structures, with up to 34 sequences in each of the most popular structures, and the distribution dramatically deviates from the null model of random population of folds in the form of Poisson distribution (blue curve).

**Figure S2.**

Distribution of mutual Hamming distances between random sequences (black curve) and within superfamilies of sequences corresponding to the three most populated wonderfolds (red, green, and blue curves).

**Figure S3.**

**a.** Dependence of the number of evolved sequences corresponding to a structure on the maximum eigenvalue ($\lambda_{max}$) of the structure's contact matrix. One can clearly see that only most designable structures – wonderfolds – form large superfamilies containing many (up to 34) dissimilar sequences. A notable property of compact 27-mers is the gap in the distribution of the eigenvalues at $\lambda_{max} \approx 2.92$, which naturally separates wonderfolds from the bulk of the structures.

**b.** Dependence of the logarithm of the average number of evolved sequences k within a bin $\lambda_{max,i} < \lambda_{max} < \lambda_{max,i} + \Delta\lambda$, with $\lambda_{max,i}$ corresponding to 50 bins spanning the range from 2.52 to 2.96. All 103346 compact structures (including those that were not found in evolution simulations, i.e. having k=0) are considered here  A linear approximation (black line) is shown, correlation coefficient R=0.97.

**Figure S4.**

Distribution of melting temperatures of 20000 proteins with random sequences (black histogram) and 537 evolved sequences that fold into 20 most populated wonderfolds (red histogram). Evolved sequences melt at significantly higher temperatures than random ones.

**Figure S5.**

Correlation between the maximum eigenvalue $\lambda_{max}$ of the structures and the melting temperature $T_m$. Structures with evolved sequences (green), and especially the top 20 wonderfolds (red) correspond statistically to higher $\lambda_{max}$ and $T_m$ than random ones (black). For random sequences, the average $\lambda_{max}$ is 2.71 (standard deviation $\sigma=0.069$), and average $T_m$ is 0.259 ($\sigma=0.144$); for all evolved structures and top 20 wonderfolds the average $\lambda_{max}$ is 2.846 ($\sigma=0.049$) and 2.92 ($\sigma=0.023$), and average $T_m$ equals 0.725 ($\sigma=0.358$) and 0.91 ($\sigma=0.35$), respectively.

**Figure S6.**

Distribution of melting temperatures of 1000 sequences designed into the top 20 wonderfolds (short design runs of 2000 MC steps, red histogram; long design runs of 200000 steps, orange histogram), and into 20 randomly selected structures (short design runs for the black histogram and long design runs for the blue histogram, see in the text). In all cases, sequences designed into wonderfolds yield higher melting temperatures.

**Figure S1.**

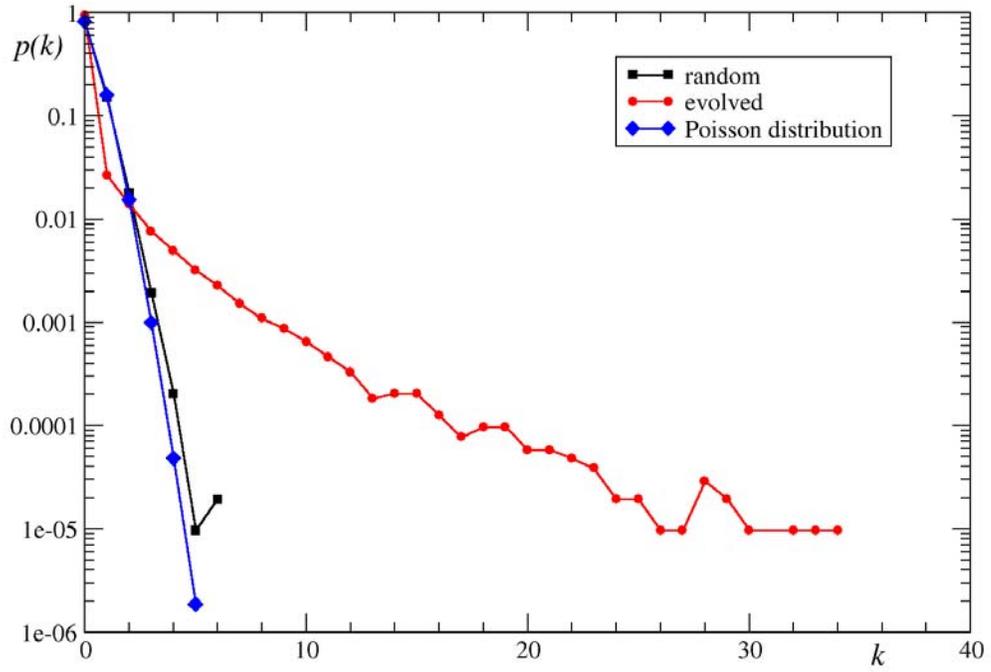

**Figure S2.**

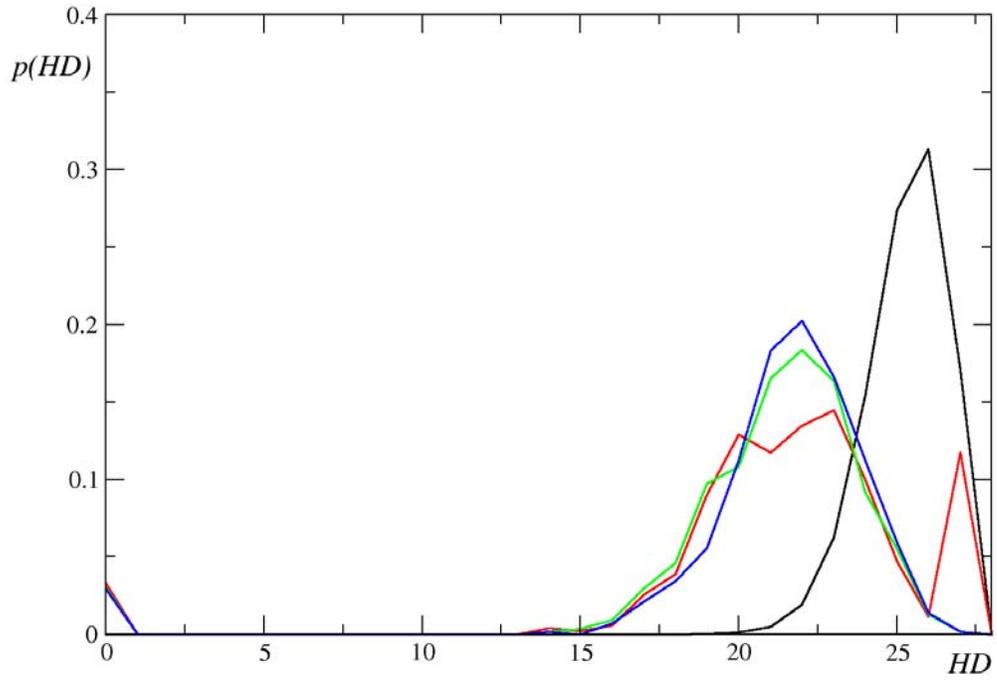

**Figure S3a.**

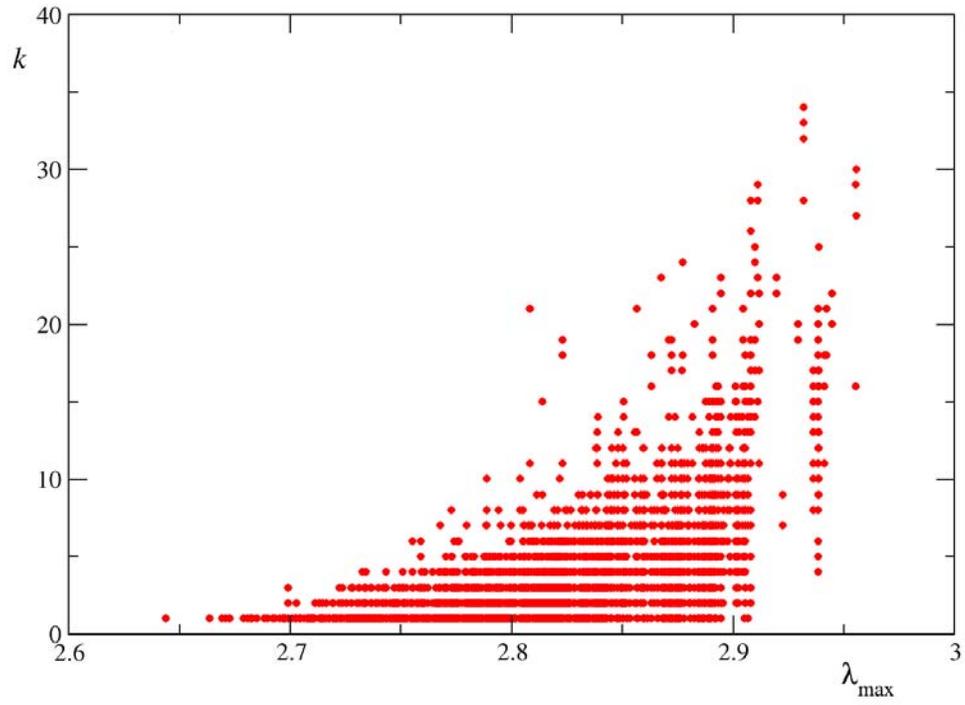

**Figure S3b.**

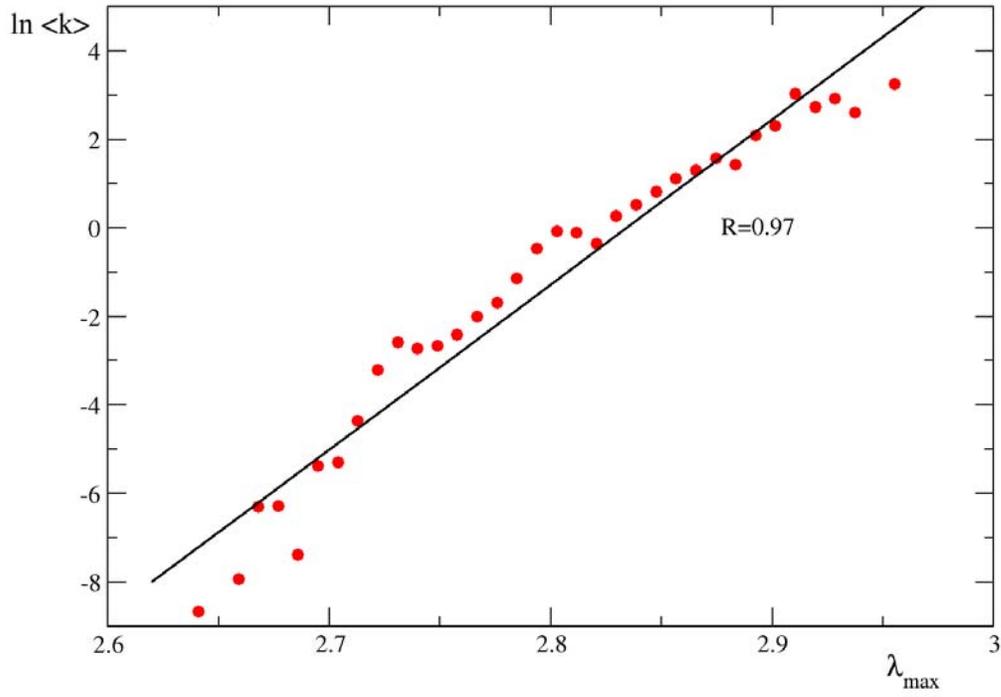

**Figure S4.**

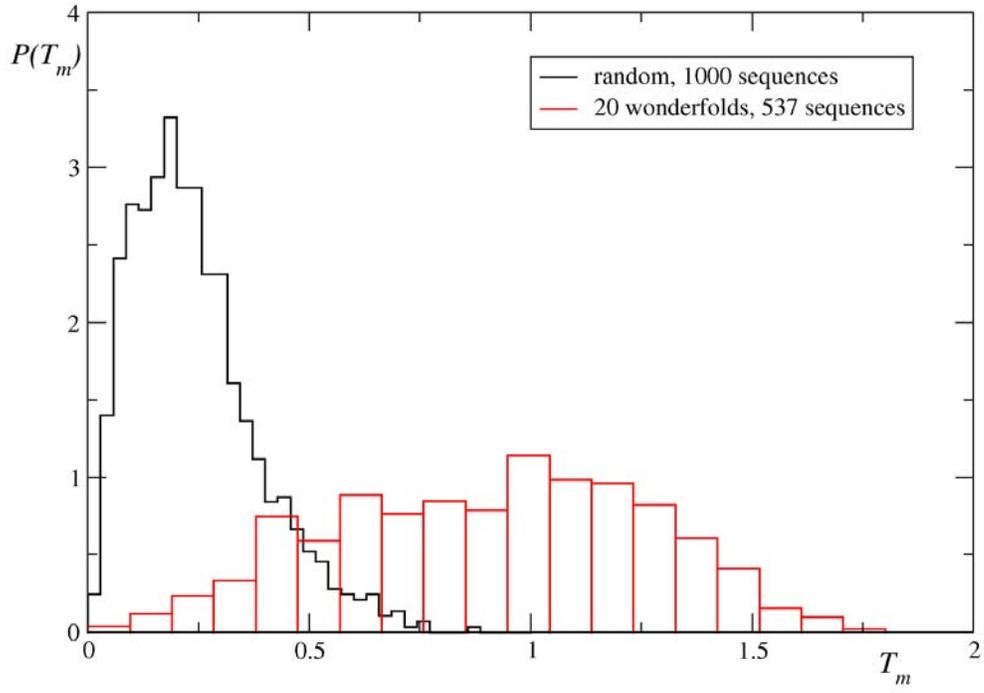

**Figure S5.**

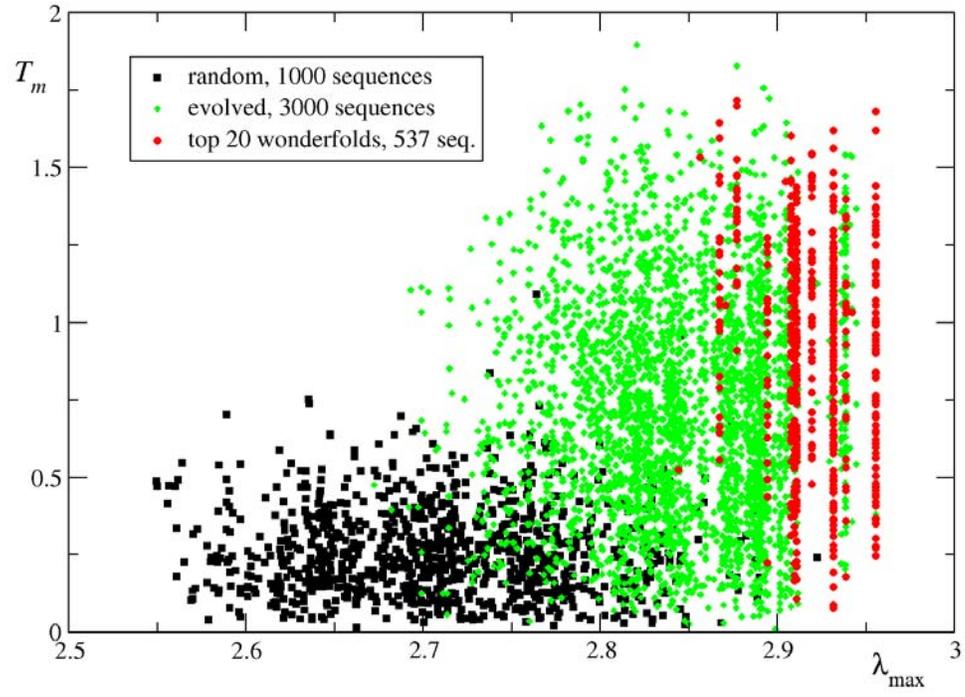

**Figure S6.**

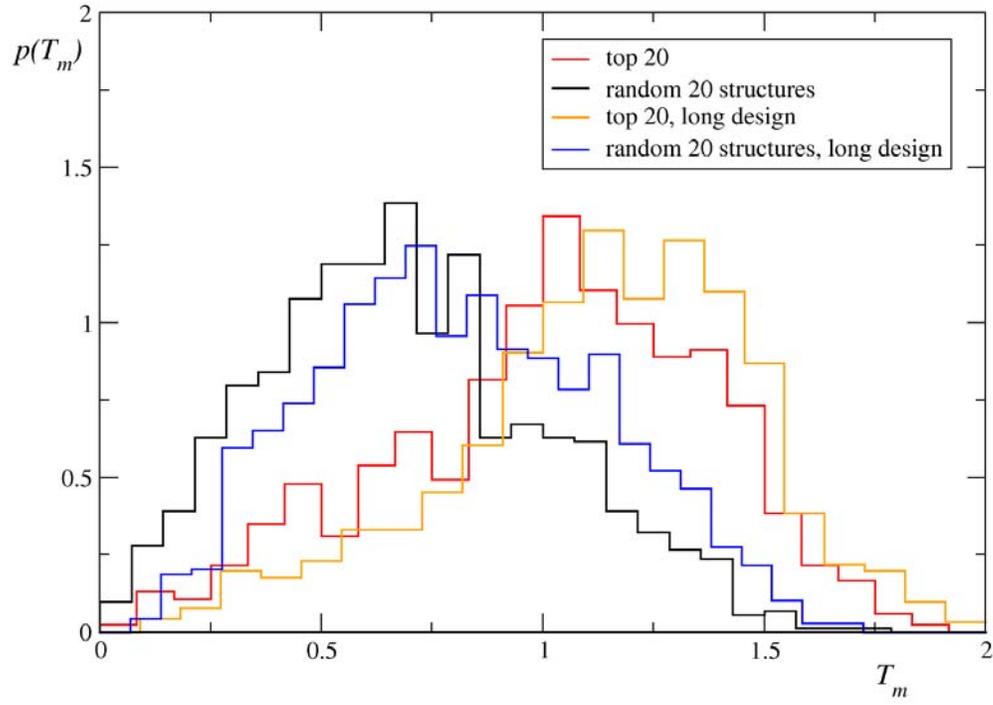